# A Density Matrix-based Algorithm for Solving Eigenvalue Problems

Eric Polizzi*

*Department of Electrical and Computer Engineering, University of Massachusetts, Amherst*
(Dated: January 17, 2009)

A new numerical algorithm for solving the symmetric eigenvalue problem is presented. The technique deviates fundamentally from the traditional Krylov subspace iteration based techniques (Arnoldi and Lanczos algorithms) or other Davidson-Jacobi techniques, and takes its inspiration from the contour integration and density matrix representation in quantum mechanics. It will be shown that this new algorithm - named FEAST - exhibits high efficiency, robustness, accuracy and scalability on parallel architectures. Examples from electronic structure calculations of Carbon nanotubes (CNT) are presented, and numerical performances and capabilities are discussed.

## I. INTRODUCTION

The generalized eigenvalue problem, that is the determination of nontrivial solutions $(\lambda, x)$ of $\mathbf{A}x = \lambda \mathbf{B}x$ with $\mathbf{A}$ and $\mathbf{B}$ square matrices, is a central topic in numerical linear algebra and arises from a wide range of applications in sciences and engineering. In electronic structure calculations, in particular, the eigenvalue problem is one of the most challenging applied numerical process - also called diagonalization procedure or spectral decomposition. In these calculations, the electron density can be formally calculated by summation of the amplitude square of the wave functions $\Psi_m$ solution of the Schrödinger-like eigenvalue problem $\mathbf{H}\Psi_m = E_m \mathbf{S}\Psi_m$ with different discrete energies $E_m$ (where $\mathbf{H}$ represents the Hamiltonian Hermitian matrix and $\mathbf{S}$ is a symmetric positive matrix obtained using non-orthogonal basis functions). This procedure can be quite computationally challenging for large-scale simulations of systems containing more than a hundred of atoms and/or where a large number of eigenpairs $(E_m, \Psi_m)$ are needed. Progress in electronic structure calculations as for other large-scale modern applications, are then much likely dependent on advances in diagonalization methods.

In the past decades, the eigenvalue problem has led to many challenging numerical questions and a central problem [1]: how can we compute eigenvalues and eigenvectors in an efficient manner and how accurate are they? Powerful tools have then been developed from Jacobi method and power iterations, to iterative Krylov subspace techniques (including Arnoldi, and Lanczos methods), or other Davidson-Jacobi techniques [2]. Traditional numerical algorithms and library packages are yet facing new challenges for addressing the current large-scale simulations needs for ever higher level of efficiency, accuracy and scalability in modern parallel architectures.

This article presents a new robust and scalable algorithm design for solving the eigenvalue problem- named FEAST- which deviates fundamentally from the traditional techniques above, and takes its inspiration from the density matrix representation and contour integration in quantum mechanics. Section II summarizes the electronic structure and contour integration problems which have motivated the development of the new algorithm. The FEAST algorithm is then described in detail in Section III, and numerical examples and performance results are presented in Section IV. Finally, Section V presents some discussions regarding the efficiency, robustness and scalability of the algorithm.

## II. THE CONTOUR INTEGRATION TECHNIQUE IN ELECTRONIC STRUCTURE CALCULATIONS

Although new fast sparse solvers have allowed considerable time saving for obtaining the eigenpairs $(E_m, \Psi_m)$ in electronic structure calculations, such as the Rayleigh-quotient multigrid [3] developed for the MIKA package, or the parallel Chebyshev subspace iteration technique developed for the PARSEC package [4, 5], these calculations are still considered computationally extremely challenging and linear scalability is not easily achievable.

An alternative approach to the Schrödinger picture for obtaining the electron density consists in performing a contour integration of the diagonal elements of the Green's function matrix $\mathbf{G}(Z) = (Z\mathbf{S} - \mathbf{H})^{-1}$ over the complex energy space [6]. At zero temperature, the resulting expression for the electron density in real-space is:

$$n(\mathbf{r}) = -\frac{1}{2\pi i} \int_{\mathcal{C}} dZ\ G(\mathbf{r}, \mathbf{r}, Z) = \sum_{m} |\Psi_m(\mathbf{r})|^2, \quad (1)$$

where the complex contour $\mathcal{C}$ includes all the eigenvalues $E_m$ below the Fermi level $E_F$, and where the spin factor is not considered. It should be noted that at non zero temperatures, this expression would also include the contribution of the residues of all poles of the Fermi-Dirac distribution function on the imaginary axis at the position of the Fermi level [7]. For transport problems and open systems, in turn, the contour integration is often used to compute the equilibrium part of the electron density [8] where self-energy boundary conditions need to be included in the Hamiltonian matrix $\mathbf{H}$. The contour integration technique represents a priori an attractive alternative approach to the traditional eigenvalue problem for computing the electron density since the number of Green's function to be calculated -typically $\sim O(10)$ using a Gauss quadrature procedure- is independent of the size of the system. In particular, an efficient linear scaling strategy CMB (CMB stands for Contour integration - Mode approach - Banded system solver), has been proposed in [9, 10] for simulating nanowire-type structures within a real-space mesh framework while overcoming the impossible numerical task of inverting a large size matrix at each point of the contour. For arbitrary systems (i.e. beyond nanowire structures), however, there are no numerical advantages of abandoning the traditional eigenvalue problem in favor of the contour integration technique for computing the electron density. In addition, it is clear from equation (1) that the contour integration technique does not provide a natural route for obtaining the individual eigenvectors but rather the summation of their amplitudes square. In the following section, a new numerical algorithm design FEAST is proposed for obtaining directly the eigenpairs solutions using the density matrix representation and a numerically efficient contour integration technique.

---

*URL: http://www.ecs.umass.edu/ece/polizzi

## III. FEAST

### A. Introduction

In this section, a new algorithm is presented for solving generalized eigenvalue problems of this form

$$\mathbf{A}\mathbf{x} = \lambda\mathbf{B}\mathbf{x}, \qquad (2)$$

within a given interval $[\lambda_{\min}, \lambda_{\max}]$, where $\mathbf{A}$ is real symmetric or Hermitian and $\mathbf{B}$ is a symmetric positive definite (s.p.d). One common way to accelerate the convergence rate of traditional iterative techniques consists in performing a factorization of the Green's function $\mathbf{G}(\sigma) = (\sigma\mathbf{B} - \mathbf{A})^{-1}$ for some reasonable shift $\sigma$ close to the eigenvalues in the search interval and which leads to solving linear systems (i.e. shifting strategy). More recently, Sakurai et al. [11, 12] have proposed a root finding technique which consists of a contour integration of a projected Laurent series-type decomposition of the Green's function. In principle, a set of complex moments can be obtained by solving few linear systems along the contour, which can generate an identical subspace to the one spanned by the eigenvectors present inside the contour. In practice, however, robustness and accuracy are not easily achievable. In our approach, we avoid decomposing directly the Green's function and perform instead an exact mathematical factorization of its contour integration - which represents the reduced density matrix $\rho$ in quantum mechanics. One can show that this factorization can be expressed in terms of the eigenvectors present inside the contour as follows:

$$\rho = -\frac{1}{2\pi\imath}\int_{\mathcal{C}} d\mathrm{Z}\ \mathbf{G}(\mathrm{Z}) == \sum_{m=1}^{\mathrm{M}} |\mathbf{x}_m\rangle\langle\mathbf{x}_m|. \qquad (3)$$

In matrix notations the second term of the equation reads $\mathbf{X}\mathbf{X}^\mathbf{T}$ where $\mathbf{X}_{N\times M} = \{\mathbf{x_1}, \mathbf{x_2}, ..\mathbf{x_M}\}$ (M being the number of eigenvalue inside the contour and N the size of $\mathbf{G}$). It should be noted that the diagonal elements of $\rho$ represent the electron density in quantum mechanics (1) discussed in Section II.

Postmultiplying $\rho$ by a set of M linearly independent random vectors $\mathbf{Y}_{N\times M} = \{\mathbf{y_1}, \mathbf{y_2}, ..\mathbf{y_M}\}$, the first expression in (3) leads to a new set of M independent vectors $\mathbf{Q}_{N\times M} = \{\mathbf{q_1}, \mathbf{q_2}, ..\mathbf{q_M}\}$ obtained by solving linear systems along the contour

$$\mathbf{Q}_{N\times M} = -\frac{1}{2\pi\imath}\int_{\mathcal{C}} d\mathrm{Z}\ \mathbf{G}(\mathrm{Z})\mathbf{Y}_{N\times M}, \qquad (4)$$

while the second expression in (3), implies these vectors $\mathbf{Q}$ can be formally generated by the eigenfunctions $\mathbf{X}$ inside the contour

$$\mathbf{Q}_{N\times M} = \mathbf{X}_{N\times M}\mathbf{W}_{M\times M} \quad \text{with} \quad \mathrm{W}_{i,j} = \mathbf{x_i^T}\mathbf{y_j}. \qquad (5)$$

In other words, each $\mathbf{Q}$ column vector obtained in (4) represents a different linear combination of unknown basis functions $\mathbf{X}$ in (5). Using a Rayleigh-Ritz procedure, the problem (2) is now equivalent to computing the eigenpairs $(\epsilon_\mathbf{m}, \mathbf{\Phi_m})$ of the following reduced generalized eigenvalue problem of size M:

$$\mathbf{A_Q}\mathbf{\Phi} = \epsilon\mathbf{B_Q}\mathbf{\Phi} \qquad (6)$$
$$\text{with} \quad \mathbf{A_Q} = \mathbf{Q^T A Q} \quad \text{and} \quad \mathbf{B_Q} = \mathbf{Q^T B Q}. \qquad (7)$$

The Ritz values and vectors are then given by:

$$\lambda_m = \epsilon_m, \quad m = 1,\ldots,\mathrm{M} \qquad (8)$$
$$\mathbf{X}_{N\times M} = \mathbf{Q}_{N\times M}\mathbf{\Phi}_{M\times M} \qquad (9)$$

where $\mathbf{\Phi}_{M\times M} = \{\mathbf{\Phi_1}, \mathbf{\Phi_2}, ..\mathbf{\Phi_M}\}$. One can show that the obtained eigenvectors $\mathbf{X}$ are naturally $\mathbf{B}$-orthonormal i.e. $\mathbf{x_i^T}\mathbf{B}\mathbf{x_j} = \delta_{\mathbf{i,j}}$, if the eigenvectors of the reduced problem (6) are $\mathbf{B_Q}$-orthonormal i.e. $\mathbf{\Phi_i^T}\mathbf{B_Q}\mathbf{\Phi_j} = \delta_{\mathbf{i,j}}$.

### B. Practical Considerations and Pseudocode

In practice, the vectors $\mathbf{Q}$ are computed by performing a numerical integration of each vectors $\mathbf{G}(\mathrm{Z})\mathbf{Y}$ (4) along the complex contour $\mathcal{C}$. Let us consider a circle centered in the middle of the search interval $[\lambda_{\min}, \lambda_{\max}]$, it should be noted that the expression of the contour integration can be further simplified since $\mathbf{G}(\bar{Z}) = \mathbf{G}^\dagger(Z)$. Denoting $\mathcal{C}^+$ the positive half circle of the complex contour, it comes if $\mathbf{A}$ is Hermitian:

$$\rho = -\frac{1}{2\pi\imath}\int_{\mathcal{C}^+} dZ\, \{\mathbf{G}(Z) - \mathbf{G}^\dagger(Z)\}, \qquad (10)$$

and if $\mathbf{A}$ is real symmetric:

$$\rho = -\frac{1}{\pi}\int_{\mathcal{C}^+} dZ\, \Im\{\mathbf{G}(Z)\}, \qquad (11)$$

where $\Im\{\}$ stands for the imaginary part. Using a $N_e$-point Gauss-Legendre quadrature on the positive half circle $\mathcal{C}^+$ (see Figure 1), with $x_e$ the $e^{th}$ Gauss node associated with the weight $\omega_e$, one obtains if $\mathbf{A}$ is Hermitian and $\mathbf{Y}, \mathbf{Q} \in \mathbb{C}^{N\times M}$:

$$\mathbf{Q} = -\sum_{e=1}^{N_e} \frac{1}{4}\omega_e r\big(\exp(\imath\theta_e)\,\mathbf{G}(Z_e) + \exp(-\imath\theta_e)\,\mathbf{G}^\dagger(Z_e)\big)\mathbf{Y}, \qquad (12)$$

with

$$r = \frac{\lambda_{\max} - \lambda_{\min}}{2}, \quad \theta_e = -(\pi/2)(x_e - 1), \qquad (13)$$

$$Z_e = \frac{\lambda_{\max} + \lambda_{\min}}{2} + r\exp(\imath\theta_e). \qquad (14)$$

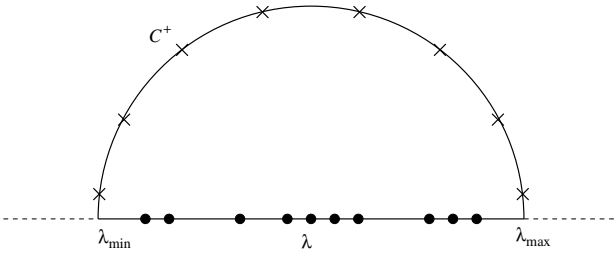

FIG. 1: Schematic representation of the complex contour integral defined by the positive half circle $\mathcal{C}^+$. In practice, the vectors $\mathbf{Q}$ are computed via a numerical integration (e.g. Gauss-Legendre quadrature) where only few linear systems $\mathbf{G}(Z)\mathbf{Y}$ needs to be solved at specific points $Z_e$ along the contour.

If $\mathbf{A}$ is real symmetric, $\mathbf{Y}, \mathbf{Q} \in \mathbb{R}^{N\times M}$ and one can use:

$$\mathbf{Q} = -\sum_{e=1}^{N_e} \frac{1}{2}\omega_e \Re\{r\exp(\imath\theta_e)\,\mathbf{G}(Z_e)\mathbf{Y}\}, \qquad (15)$$

where $\Re\{\}$ stands for the real part.

In order to reduce the numerical quadrature error of the contour integral, one may consider the two following improvements:
**(i)** Performing outer-iterative refinement steps. Once the eigenvectors $\mathbf{X}$ are obtained (9), a new set of initial guess vectors $\mathbf{Y} = \mathbf{BX}$ can be used. Postmultiplying the density matrix (3) by $\mathbf{Y}$, one now obtains from (5) that $\mathbf{Q}$ converges to $\mathbf{X}$ since $\mathbf{X}^T\mathbf{BX} = \mathbf{I}$ (i.e. $W_{i,j} = \delta_{i,j}$ and then $\rho\mathbf{BX} = \mathbf{X}$). A fast test for convergence can be obtained by checking the trace of the eigenvalues (8).

**(ii)** Postmultiplying the density matrix (3) by $M_0$ random vectors (rather than M) where $M_0$ is greater than M. The reduced dense generalized eigenvalue problem (6) of size $M_0$ can be solved using standard eigenvalue LAPACK routines [13]. Since we do not perform the orthogonalization of the vectors $\mathbf{Q}$, one has to make sure that $\mathbf{B}_{\mathbf{Q}_{M_0},M_0}$ is symmetric positive definite i.e. $M_0$ does not exceed an upper limit which can easily be obtained a posteriori.

---

**1-** Select $M_0 > M$ random vectors $\mathbf{Y}_{N\times M_0} \in \mathbb{R}^{N\times M_0}$
**2-** Set $\mathbf{Q} = 0$ with $\mathbf{Q} \in \mathbb{R}^{N\times M_0}$; $r = (\lambda_{\max} - \lambda_{\min})/2$;
    For $e = 1,\ldots N_e$
        compute $\theta_e = -(\pi/2)(x_e - 1)$,
        compute $Z_e = (\lambda_{\max} + \lambda_{\min})/2 + r\exp(\imath\theta_e)$,
        solve $(Z_e\mathbf{B} - \mathbf{A})\mathbf{Q_e} = \mathbf{Y}$ to obtain $\mathbf{Q_e} \in \mathbb{C}^{N\times M_0}$
        compute $\mathbf{Q} = \mathbf{Q} - (\omega_e/2)\Re\{r\exp(\imath\theta_e)\,\mathbf{Q_e}\}$
    End
**3-** Form $\mathbf{A}_{\mathbf{Q}_{M_0\times M_0}} = \mathbf{Q}^T\mathbf{A}\mathbf{Q}$ and $\mathbf{B}_{\mathbf{Q}_{M_0\times M_0}} = \mathbf{Q}^T\mathbf{B}\mathbf{Q}$
**4-** Solve $\mathbf{A_Q}\mathbf{\Phi} = \epsilon\mathbf{B_Q}\mathbf{\Phi}$ to obtain the $M_0$ eigenvalue $\epsilon_m$,
    and eigenvectors $\mathbf{\Phi}_{M_0\times M_0} \in \mathbb{R}^{M_0\times M_0}$
**5-** Set $\lambda_m = \epsilon_m$ and compute $\mathbf{X}_{N\times M_0} = \mathbf{Q}_{N\times M_0}\mathbf{\Phi}_{M_0\times M_0}$
    If $\lambda_m \in [\lambda_{\min}, \lambda_{\max}]$, $\lambda_m$ is an eigenvalue solution
    and its eigenvector is $\mathbf{X_m}$ (the $m^{th}$ column of $\mathbf{X}$).
**6-** Check convergence for the trace of the eigenvalues $\lambda_m$
    If iterative refinement is needed, compute $\mathbf{Y} = \mathbf{BX}$
    and go back to step 2

FIG. 2: FEAST pseudocode (sequential version) for solving the generalized eigenvalue problem $\mathbf{Ax} = \lambda\mathbf{Bx}$, where $\mathbf{A}$ is real symmetric and $\mathbf{B}$ is s.p.d., and obtaining all the M eigenpairs within a given interval $[\lambda_{\min}, \lambda_{\max}]$. The numerical integration is performed using $N_e$-point Gauss-Legendre quadrature with $x_e$ the $e^{th}$ Gauss node associated with the weight $\omega_e$. For the case $N_e = 8$, one can use:
$(x_1, \omega_1) = (0.183434642495649, 0.362683783378361)$,
$(x_3, \omega_3) = (0.525532409916328, 0.313706645877887)$,
$(x_5, \omega_5) = (0.796666477413626, 0.222381034453374)$,
$(x_7, \omega_7) = (0.960289856497536, 0.101228536290376)$,
and $(x_{2i}, \omega_{2i})_{i=1,\ldots,4} = (-x_{2i-1}, \omega_{2i-1})$

The performances of the basic FEAST algorithm will then depend on a trade off between the choices of the number of Gauss quadrature points $N_e$, the size of the subspace $M_0$, and the number of outer refinement loops. So far, using $M_0 \geq 1.5M$, $N_e = 8$, and with at most 2 refinement loops, we have consistently obtained a relative residual equal or smaller than $10^{-10}$ seeking up to 1000 eigenpairs for a variety of problems. The basic pseudocode for the FEAST algorithm is given in Figure 2 in the case of $\mathbf{A}$ real symmetric. In the case of $\mathbf{A}$ complex Hermitian, we note the following changes:

$\mathbf{Y}, \mathbf{Q} \in \mathbb{C}^{N \times M_0}$, $\mathbf{\Phi} \in \mathbb{C}^{M_0 \times M_0}$, and the construction of the vectors $\mathbf{Q}$ in step-**2** of the pseudocode must be modified to satisfy (12).

## IV. NUMERICAL EXPERIMENTS

In this section, we propose to demonstrate the numerical stability, robustness and scalability of the FEAST algorithm using three examples derived from electronic structure calculations of Carbon nanotube (CNT).

### A. Example I

Let us first consider a family of eigenvalue problems, Test-CNT, obtained using a 2D FEM discretization of the DFT/Kohn-Sham equations at a given cross-section of a (13,0) CNT – the 2D atomistic potential is derived from the mode approach used in the CMB strategy for solving the full 3D problem presented in [9]. In Test-CNT, $\mathbf{A}$ is real symmetric, and $\mathbf{B}$ is s.p.d., the size of both matrices is $N = 12,450$ and their sparsity pattern is identical with a number of non-zero elements $nnz = 86,808$.

In Table I, we report the times and relative residual obtained by the public domain eigenvalue solver package ARPACK [14] (using the shift-invert strategy) and the FEAST algorithm presented in Figure 2 for solving the Test-CNT example seeking up to M = 800 (lowest) eigenpairs. The inner linear systems in ARPACK and FEAST are solved using the shared-memory parallel direct solver PARDISO [15]. It should be noted, however, that FEAST benefits more than ARPACK from the PARDISO solver, as the inner linear systems have multiple right-hand sides. Although both algorithms could benefit from a parallel distributed implementation (e.g. using the P_ARPACK package), the simulation runs are here restricted to a given node of a 8-cores Intel Clovertown system (16Gb,2.66GHz) where the linear systems in FEAST are factorized and solved one after another. The performances of ARPACK and FEAST can also depend on fine tunings parameters such as the choices of the size of the subspace $M_0$ ($M_0 = 1.5M$ here for both algorithms), the inner systems solvers, the number of contour points $N_e$ for FEAST, or the stopping criteria for obtaining the residual. The simulation results in this section are then not intended to compare quantitatively the two solvers but rather to point out the potentialities of the FEAST algorithm.

In our experiments, the convergence criteria on the relative residual for FEAST is obtained when the relative error on the trace of the eigenvalues $\sum_m \lambda_m$ in the search interval is smaller or equal to $10^{-13}$. Table II shows the variation of the relative error on the trace with the number of outer-iterative refinement for FEAST. These results demonstrate that only 2 to 3 refinement loops are necessary to obtain the small relative residuals for the different cases reported in Table I. It should be noted that only one loop is necessary to obtain the eigenvalues with an accuracy of $\sim 10^{-5}$ or below.

| TEST-CNT | ARPACK | | FEAST | |
| N = 12,450 | Time(s) | Resid. | Time(s) | Resid. |
| --- | --- | --- | --- | --- |
| M = 100 | 12.2 | $2.0 * 10^{-11}$ | 7.8 | $4.5 * 10^{-10}$ |
| M = 200 | 31 | $2.0 * 10^{-11}$ | 14 | $5.5 * 10^{-10}$ |
| M = 400 | 86 | $1.4 * 10^{-11}$ | 21 | $1.8 * 10^{-10}$ |
| M = 800 | 213 | $4.5 * 10^{-9}$ | 58 | $3.4 * 10^{-11}$ |

TABLE I: Simulations times and relative residual $\max_i(\|\mathbf{A}\mathbf{x_i} - \lambda_i \mathbf{B}\mathbf{x_i}\|_1 / \|\mathbf{A}\mathbf{x_i}\|_1)$, obtained by the solver ARPACK and FEAST on the TEST-CNT system seeking M (lowest) eigenpairs for different search intervals. The simulations are performed on a Intel Clovertown (8cores, 1 node, 2.66GHz, 16Gb). The shift-strategy has been used in ARPACK to accelerate the convergence (the regular mode would give $\sim 300s$ for M = 100). The inner linear systems in ARPACK and FEAST are both solved using the direct parallel solver PARDISO [15] on 8 cores. Finally, the size of the subspace has been chosen to be $M_0 = 1.5M$ for both algorithms, and the number of contour points for FEAST is fixed at $N_e = 8$.

| TEST-CNT | Relative error on the Trace | | |
| N = 12,450 | 1st loop | 2nd loop | 3rd loop |
| --- | --- | --- | --- |
| M = 100 | $3.0 * 10^{-6}$ | $2.9 * 10^{-12}$ | $1.0 * 10^{-15}$ |
| M = 200 | $1.8 * 10^{-5}$ | $4.8 * 10^{-12}$ | $2.1 * 10^{-14}$ |
| M = 400 | $2.4 * 10^{-8}$ | $3.2 * 10^{-16}$ | |
| M = 800 | $1.8 * 10^{-9}$ | $4.3 * 10^{-16}$ | |

TABLE II: Variation of the relative error on the trace of the eigenvalues $\sum_m \lambda_m$ for different search intervals with the number of iterative refinement loops. The convergence criteria is set to $10^{-13}$ where the final relative residual on the eigenpairs is reported in Table I.

The simulation results in Table I demonstrate very good scalability for FEAST while the search interval keeps increasing but the number of contour points $N_e$ stays identical (i.e. the number of numerical operations stays the same for a given loop of FEAST with a fixed $N_e = 8$ linear systems to solve). In addition, from Table III, one can see how the robustness of the FEAST algorithm is affected while the number of contour points $N_e$ changes. In particular, $N_e = 4$ points along the contour did suffice to capture M = 100 eigenpairs with a relatively small residual (decreasing the simulation time reported in Table I for this case), while the case $N_e = 16$ points generated a residual smaller than the one obtained by ARPACK (using $M_0 = 1.5M$).

### B. Example II

In another set of numerical experiments, we intend to demonstrate the robustness of FEAST in capturing the multiplicity of the eigenvalues. We propose to create artificially new TEST-CNT systems called $k(N, M)$ where the matrices $\mathbf{A}$ and $\mathbf{B}$ are repeated $k$ times along the main diagonal (the new system matrix is block diagonal with $k$ blocks). Physically, these systems can describe the cross section of a bundle composed by $k$ CNTs, where we do not consider the interactions between the different tubes such that each eigenvalue is now $k$ times degenerate. If we keep the same search interval used to obtain M = 100 eigenpairs for $k = 1$ (where the size of the matrices $\mathbf{A}$ and $\mathbf{B}$ is N), $100k$ eigenpairs must now be found for $k \geq 1$, where each one of them have the multiplicity $k$. In Table IV, we report the simulation times and relative residuals obtained using ARPACK and FEAST on these $k(N, M)$ TEST-CNT systems. For the case $8(N, M)$, for example, the size of the new system matrix is 99, 600 and the first 100 eigenvalues have all the multiplicity 8 (so 800 eigenpairs are found in total). The simulation results show linear scalability performances with the size of the system and the number of eigenpairs. In contrast to ARPACK where the number of matrix-vector multiplications and linear system solves would keep increasing with $k$, the number of operations in FEAST stays the same for all these cases. The scalability of the algorithm depends then mainly on the scalability of the linear system solver.

| TEST-CNT | FEAST | | |
|---|---|---|---|
| M = 100 | Time(s) | Resid. | # loops |
| $N_e = 4$ | 7.0 | $8.3 * 10^{-8}$ | 6 |
| $N_e = 8$ | 7.8 | $4.5 * 10^{-10}$ | 4 |
| $N_e = 16$ | 10.2 | $3.4 * 10^{-12}$ | 3 |

TABLE III: Performance results obtained by FEAST seeking M = 100 eigenpairs for different values of $N_e$. The convergence is obtained when the error on the trace is equal or smaller to $10^{-13}$.

| TEST-CNT | ARPACK | | FEAST | |
|---|---|---|---|---|
| N = 12, 450 M = 100 | Time(s) | Resid. | Time(s) | Resid. |
| (N,M) | 12.2 | $2.0 * 10^{-11}$ | 7.8 | $4.5 * 10^{-10}$ |
| 2(N,M) | 85 | $3.5 * 10^{-11}$ | 27 | $7.7 * 10^{-10}$ |
| 4(N,M) | 668 | $4.6 * 10^{-11}$ | 109 | $8.8 * 10^{-10}$ |
| 8(N,M) | 5492 | $6.2 * 10^{-11}$ | 523 | $6.5 * 10^{-10}$ |

TABLE IV: Simulations times and relative residual $\max_i(||\mathbf{Ax_i} - \lambda_i\mathbf{Bx_i}||_1/||\mathbf{Ax_i}||_1)$, obtained by the solver ARPACK and FEAST on the $k(N, M)$ TEST-CNT systems which artificially reproduce $k$ times the original TEST-CNT system. The $kM$ (lowest) eigenpairs are found where each eigenvalue has a multiplicity of $k$.

## C. Example III

We have shown that FEAST can re-use the computed subspace as suitable initial guess for performing iterative refinements. This capability can also be of benefit to modern applications in science and engineering where it is often necessary to solve a series of eigenvalue problems that are close one another. In bandstructure calculations, in particular, many eigenvalue problems of the form $(\mathbf{A} + \mathbf{S_k})\mathbf{x_k} = \lambda_k\mathbf{Bx_k}$ need to be solved at different locations in the $\mathbf{k}$-space (i.e. for different values of $\mathbf{k}$ and where $\mathbf{S}$ is Hermitian with $\mathbf{S_0} = \mathbf{I}$). Let us consider the eigenvalue sparse system of size N = 492, 982 obtained for a (5,5) metallic CNT using our in-house DFT/real-space mesh technique framework for bandstructure calculations of nanowires-type structure [16]. In Figure 3, we propose to solve this eigenvalue problem using the same search interval for the eigenvalues $\lambda$ for different locations of $\mathbf{k}$ where the subspace computed by FEAST at the point $\mathbf{k} - \mathbf{1}$ is successively used as initial guess for the neighboring point $\mathbf{k}$. In addition, the inner linear systems in FEAST are solved using an iterative method with preconditioner where a modest relative residual of $10^{-3}$ is used (e.g. a suitable banded preconditioner can be obtained using a mode approach [9]). It should be noted that the convergence criteria for the relative error on the trace of the eigenvalues is chosen much smaller at $10^{-8}$, while the eigenvectors are expected to be obtained within the same (or a smaller) order of accuracy that the one used for the solutions of the inner systems. Figure 3 shows that 13 to 20 eigenvalues (i.e. energies) are found within the selected search interval along the different **k** points (from the **Γ** to the **X** point in the graph). Although the size of the subspace stays identical at $M_0 = 25$, after the first initial point at $\mathbf{k} = \mathbf{0}$ (Γ point in the graph) FEAST converges within only one refinement loop for almost all the other **k** points.

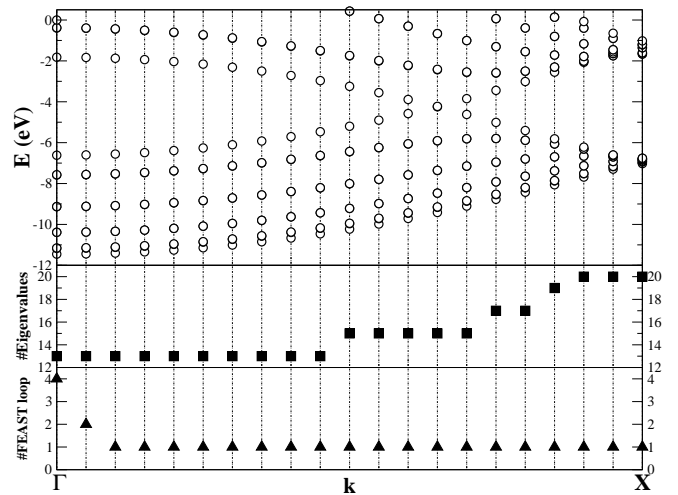

FIG. 3: Bandstructure calculations of a (5,5) metallic CNT. The eigenvalue problems are solved successively for all the $\mathbf{k}$ points (from $\mathbf{\Gamma}$ to $\mathbf{X}$), while the computed subspace of size $M_0 = 25$ at the point $\mathbf{k}$ is used as initial guess for the point $\mathbf{k} + \mathbf{1}$. The number of eigenvalues found ranges from 13 to 20, and by the third $\mathbf{k}$ point, the FEAST convergence is obtained using only one refinement loop. The convergence is obtained with the relative error on trace of the eigenvalues smaller or equal to $10^{-8}$, while the inner linear systems are solved using an iterative method with an accuracy of $10^{-3}$. The final relative residuals on the eigenpairs range from $10^{-3}$ to $10^{-5}$.

## V. DISCUSSIONS

In comparison to iterative Krylov subspace techniques, FEAST can be cast as a "direct" technique which is based on an exact mathematical derivation (3). FEAST does naturally then capture all the multiplicity and no-orthogonalization procedure is necessary (such as Gram-Schmidt orthogonalization process). As described above, the main computational tasks in FEAST consist of solving $N_e$ independent linear systems along the contour with $M_0$ right-hand-sides and a reduced dense generalized eigenvalue problem of size $M_0$. Since FEAST has the ability to re-use the basis from the previously computed subspace, an outer-iterative refinement procedure is proposed to improve the accuracy of the solutions. The capability to take advantage of suitable initial guess can also be of benefit to modern applications in sciences and engineering where it is often necessary to solve a series of eigenvalue problems that are close one another (e.g. Bandstructure calculations in Example III Section IV).

In one sense, the difficulty of solving an eigenvalue problem has been replaced by the difficulty of solving a linear system with multiple right-hand sides. For large sparse systems, this latter can be solved using either a direct system solver such as PARDISO [15] (as proposed in Section IV), or an iterative system solver with preconditioner. In turn, for banded systems or banded preconditioner, FEAST can be seen as an outer-layer for the author's SPIKE parallel system solver [17]. It should be noted that the inner linear systems arising from standard eigenvalue solvers (using the shift-strategy), need often to be solved highly accurately via direct methods. Direct system solvers, however, are not always suited for addressing large-scale modern applications because of memory requirements. In Example III of Section IV we have shown that FEAST can take advantage of iterative solvers for solving the inner linear systems with modest relative residual and obtaining the eigenvectors solution within the same order of accuracy. The resulting subspace could also be used as a very good initial guess for a one step more accurate refinement procedure (i.e. using more accurate relative residual for the inner systems).

FEAST exhibits important potentialities for parallelism at three different levels: (i) many search interval $[\lambda_{\min}, \lambda_{\max}]$ can be run independently, (ii) each linear systems can be solved simultaneously (e.g. on each node of parallel architecture where the factorization of the linear system can be done only once for all the refinement loops), (iii) the linear system solver can be parallel (e.g. within a given node as in Section IV). Depending on the parallel architecture at hand, the local memory of a given node and the properties of the matrices of the eigenvalue problems, one may preferably select one parallel option among the others, or just take advantage of a combination of those. In particular, there will be a trade off between how many search intervals to consider and how many eigenpairs FEAST can handle by intervals. For example if $M_0$ is more than few thousands, one could either (i) solve the obtained reduced system of size $M_0$ using efficient dense parallel symmetric eigenvalue solvers [18], or (ii) propose to divide the initial search interval into two or more to be processed in parallel. In addition, it should be noted that the orthonormalization step is absent from FEAST which will drastically reduce the communication overhead for performing scalar products on high-end parallel architectures (the scalar product in step-**3** in Fig. 2 has to be done only once per iterative refinement). Given the recent advances in parallel architectures and parallel linear system solvers, it is reasonable to envision using FEAST in a near future for obtaining up to millions of eigenpairs of large sparse symmetric eigenvalue problems. Finally the capabilities of FEAST could potentially be enhanced for addressing non-symmetric eigenvalue problems where the contour integration would then be performed in a given region of the complex space.

### Acknowledgments

The author wishes to acknowledge helpful discussions with Dr. Ahmed Sameh and Dr. Massimo Fischetti. This material is supported by NSF under Grant #CCF-0635196.